\documentclass[twocolumn,aps,prl,superscriptaddress,longbibliography]{revtex4-1}
\usepackage[colorlinks,bookmarks=false,citecolor=blue,linkcolor=red,urlcolor=blue]{hyperref}
\usepackage{verbatim}
\usepackage{amsmath,amssymb,bm,braket,float,mathtools}
\usepackage{graphicx,xfrac,appendix}
\usepackage[english]{babel}
\makeatletter

\usepackage{epstopdf}

\makeatother

\begin{document}
\title{From topological phase to Anderson localization in a two-dimensional quasiperiodic system}

\author{Shujie Cheng}
\affiliation{Department of Physics, Zhejiang Normal University, Jinhua 321004, China}
\author{Reza Asgari}
\affiliation{School of Physics, Institute for Research in Fundamental Sciences (IPM), Tehran 19395-5531, Iran}
\affiliation{Department of Physics, Zhejiang Normal University, Jinhua 321004, China}
\author{Gao Xianlong}
\thanks{Corresponding author: gaoxl@zjnu.edu.cn}
\affiliation{Department of Physics, Zhejiang Normal University, Jinhua 321004, China}

\date{\today}

\begin{abstract}
In this paper, the influence of the quasidisorder on a two-dimensional system is studied. 
We find that there exists a topological phase transition accompanied by a transverse 
Anderson localization. The topological properties are characterized by the band gap, 
the edge-state spectra, the transport conductance, and the Chern number. The 
localization transition is clearly demonstrated by the investigations of the partial 
inverse participation ratio, the average of level spacing ratio, and the fraction 
dimension. The results reveal the topological nature of the bulk delocalized states. 
Our work facilitates the understanding on the relationship between the topology and the 
Anderson localization in two-dimensional disordered systems.
\end{abstract}

\maketitle

\paragraph{Introduction.---}Band insulators have attracted much attention because of their peculiar topological properties. 
Two representative insulators are the Chern class \cite{Chern} and the quantum spin-Hall class \cite{spin_Hall_0,spin_Hall_1,spin_Hall_2}. 
The topological properties are reflected in that although these band insulators present the characteristics of an insulator in the bulk, 
there still exists robust conducting states, distributed at the edges of the band insulators. The Fermi levels of these edge 
modes connect the valence and conduction bands, making the insulators present metallic characteristics \cite{edge_modes}. Moreover, 
the presence or absence of the edge modes can be predicted by the Chern number ($C$). According to the Thouless-Kohmoto-Nightingale-den Nijs (TKNN)'s 
work \cite{TKNN}, $C\neq 0$ indicates the presence of the edge modes, whereas $C=0$ means the absence of the edge modes.

 \begin{figure}[htp]
 \centering
 \includegraphics[width=0.5\textwidth]{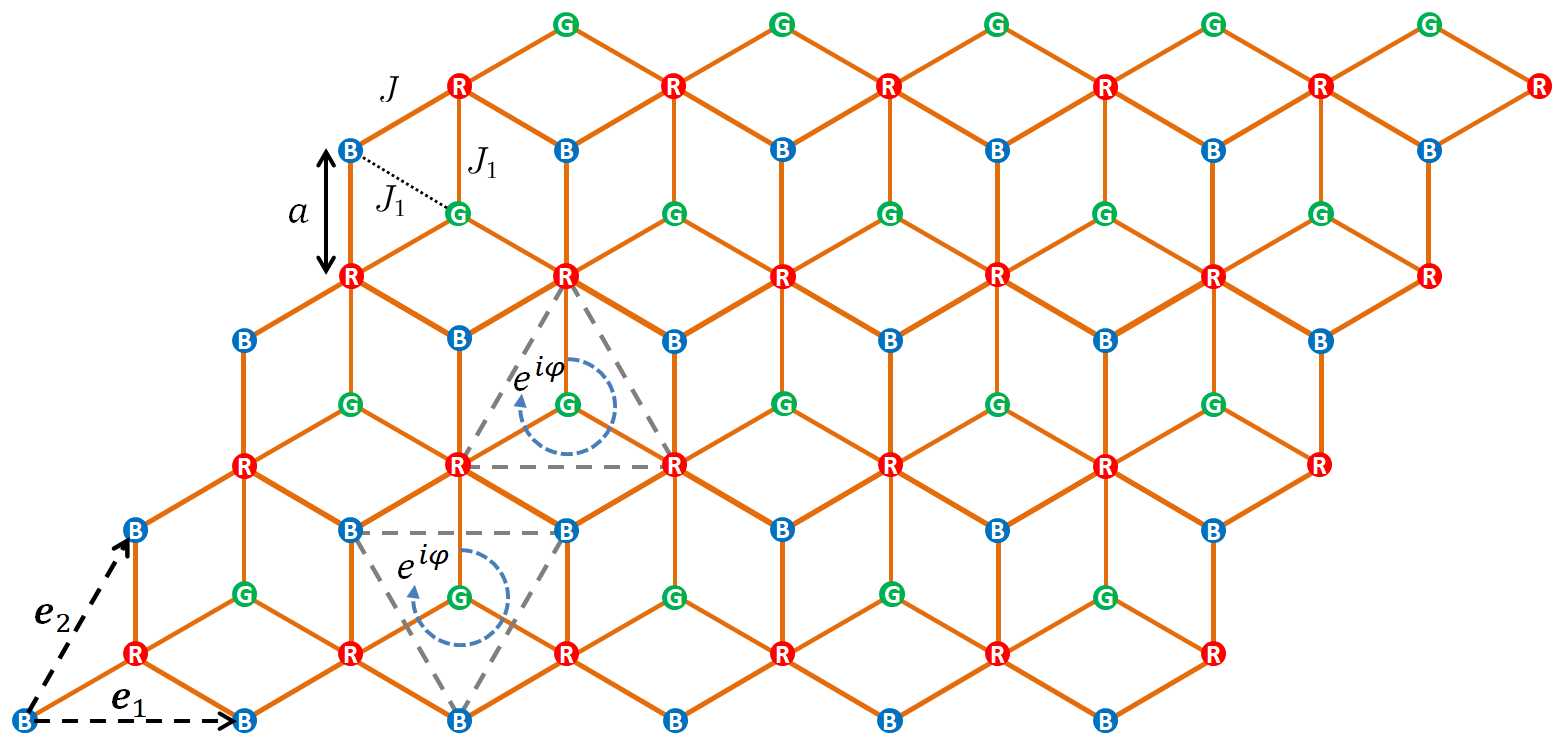}
 \caption{(Color Online) Sketch of the model. The primitive lattice vectors $\bm{e_{1}}$ and $\bm{e_{2}}$ form 
  a unit cell, which contains three types of independent sublattices, marked by $\rm{R}$, $\rm{B}$, and $\rm{G}$, respectively. 
  The hopping strength between nearest neighbor $\rm{R}$ and $\rm{B}$ sites is $J$, and the one between nearest 
  neighbor $\rm{G}$ and $\rm{R}$ (or $\rm{B}$) sites is $J_{1}$. The hopping strength between 
  two nearest neighbor $\rm{R}$ (or $\rm{B}$) sites is $J_{2}e^{i \varphi}$.}\label{f1}
 \end{figure}

The above mentioned metallic characteristics are found to be robust against the disorder, implying that 
the topological properties of the system still preserve as long as the bulk gap keeps open within a certain 
disorder strength \cite{Q.Niu}. Increasing the disorder strength will finally lead to the close of 
the bulk gap, making the system be in a topologically trivial phase \cite{Prodan_1,Prodan_2}. 
Recently, there are growing interests in studying  localization in the two-dimensional 
systems \cite{loc_0,loc_0_1,loc_0_2,loc_0_3,loc_len_1,loc_len_2,loc_1,loc_2,loc_3,loc_4,loc_5} with Anderson disorder. Particularly, 
the coexistence of nontrivial topology and Anderson localization in a quantum spin-Hall system \cite{spin_Hall_3} 
with broken time-reversal symmetry is found and the extended states are protected by the nontrivial topology \cite{spin_Hall_4}. 
It is known to us that quasidisorder is a another form of disorder and it belongs to the correlated disorder \cite{quasi_disorder} 
and the Chern insulator is one of the types of the band insulators without time-reversal symmetry. In this work, 
we are motivated to investigate whether the aforementioned conclusion about the topological nature of the bulk states 
still applies to the quasidisordered Chern insulators. After all, Ref. \cite{loc_0_3} has remarked that there may 
exist multifractal wave functions in the two-dimensional disordered systems.

\paragraph{Model.---}
Figure \ref{f1} presents the schematic diagram of the two-dimensional quasidisordered Chern insulator system. 
The primitive lattice vectors, $\bm{e_{1}}$ and $\bm{e_{2}}$, constitute the unit cell where there 
are three types of independent sublattices, marked by $\rm{R}$, $\rm{B}$, and $\rm{G}$, respectively. $a$ is the spacing between two nearest 
neighbor sites ($a=1$ in general).  The Hamiltonian ($\hat{H}$) of the system consists of two parts. One is the 
hopping terms ($\hat{H}_{1}$) and another is the quasidisordered (quasiperiodic) potentials ($\hat{H}_{2}$). The total Hamiltonian 
is $\hat{H}=\hat{H}_{1}+\hat{H}_{2}$. $\hat{H}_{1}$ reads  
\begin{equation}\label{eq1}
\begin{aligned}
\hat{H}_{1}&= \sum_{j, j'}\left(J\hat{c}^{\dag}_{\rm{R}_{j}}\hat{c}_{\rm{B}_{j'}} 
+ J_{1} \hat{c}^{\dag}_{\rm{G}_{j}}\hat{c}_{\rm{R}_{j'}} + J_{1} \hat{c}^{\dag}_{\rm{G}_{j}}\hat{c}_{\rm{B}_{j'}} \right.\\
&\left.+J_{2}e^{i \varphi}\hat{c}^{\dag}_{\rm{R}_{j}}\hat{c}_{\rm{R}_{j'}}+ J_{2}e^{i \varphi}\hat{c}^{\dag}_{\rm{B}_{j}}\hat{c}_{\rm{B}_{j'}} + h.c.\right), 
\end{aligned}
\end{equation}
where $J$ is the unit of energy and $\mathcal{S}_{m,n}=m\bm{e_{1}}+n\bm{e_{2}}$ ($m$ and $n$ are integers) 
is the position of the $\mathcal{S}$ site ($\mathcal{S} \in \{\rm{R}, \rm{B}, \rm{G}\}$). $\hat{H}_{2}$ is 
\begin{equation}\label{eq2}
\begin{aligned}
\hat{H}_{2}&=\sum_{\mathcal{S}_{m,n}}V\cos(2\pi\alpha n)\hat{c}^{\dag}_{\mathcal{S}_{m,n}}\hat{c}_{\mathcal{S}_{m,n}}, 
\end{aligned}
\end{equation}
where $V$ is the strength of the quasi-periodic potential, and $\alpha=(\sqrt{5}-1)/2$. Intuitively, this is a 
multi-parameter system. In this paper, we aim to investigate the topological properties, localization 
phase transition, and the quantum criticality cased by the quai-periodic potential. In the following, we 
take $J_{1}=0.3$, $J_{2}=0.275$, and $\varphi=\pi/2$ without loss of generality. 
 
 \begin{figure}[htp]
 \centering
 \includegraphics[width=0.5\textwidth]{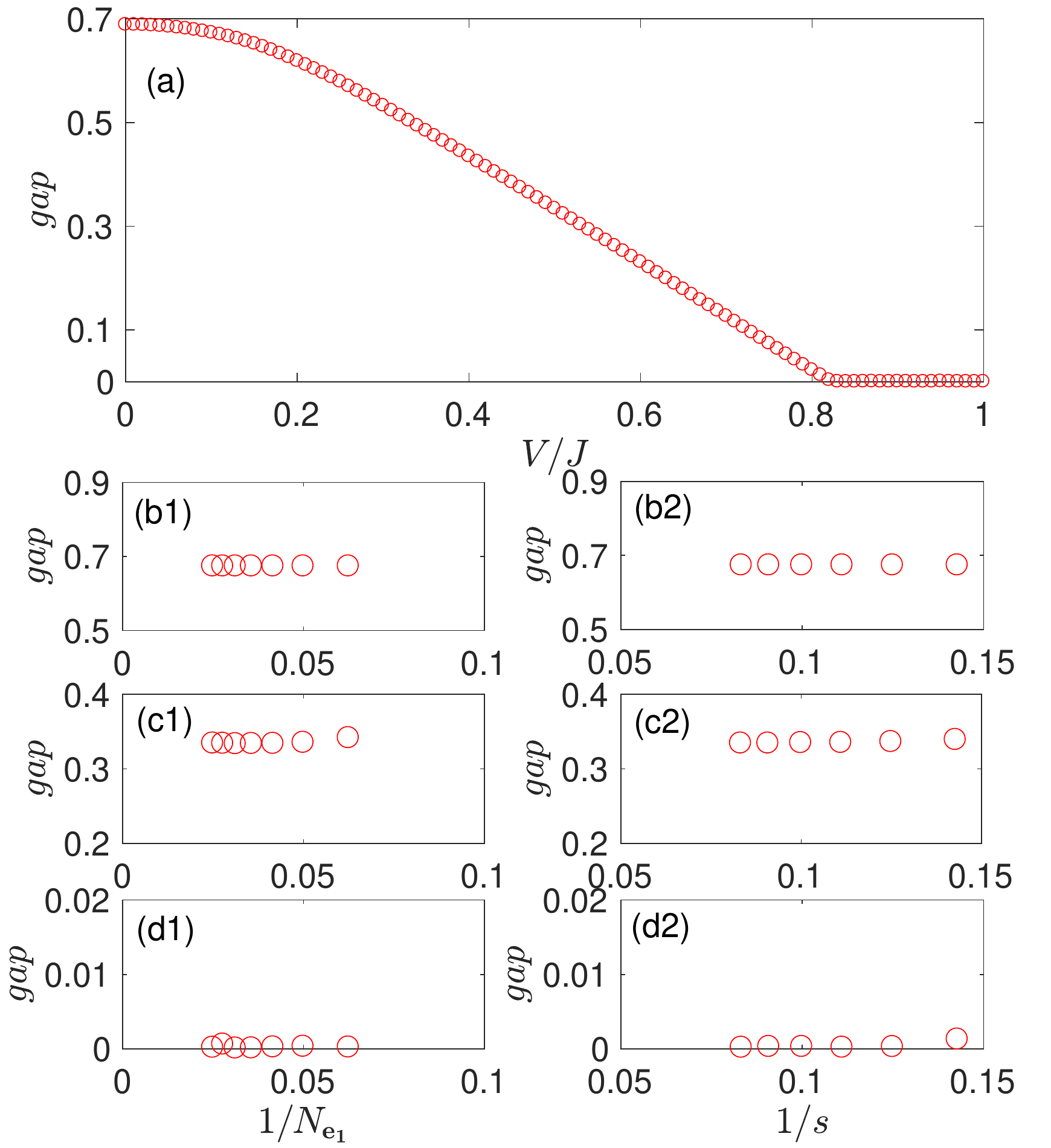}
 \caption{(Color Online) (a) The energy gap of the lower two minibands as a function of $V$ under PBC. 
 The considered system has a size of $40 \times 144$. (40 unit cells along $\bm{e_{1}}$ and $144$ 
 unit cells along $\bm{e_{2}}$). (b1)-(d2) Finite-size analysis about the energy gap with $V=0.1J$ in (b1) 
 and $(b2)$, $V=0.5J$ in (c1) and (c2), and $V=J$ in (d1) and (d2). To obtain (b1), (c1), and (d1), we keep 
 $N_{\bm{e_{2}}}$ invariant and change $N_{\bm{e_{1}}}$. While calculating (b2), (c2), and (d2), we only 
 change $N_{\bm{e_{2}}}$, and make $N_{\bm{e_{2}}}$ be equal to the $s$-th Fibonacci number. Obviously, 
 the energy gaps almost remain constants as the size of the system increases.}\label{f2}
 \end{figure}

\paragraph{Topological properties.---} Considering a system with the size $N_{\bm{e_{1}}} \times N_{\bm{e_{2}}}=40 
\times 144$ ($N_{\bm{e_{1}}}$ is the number of unit cells along $\bm{e_{1}}$ (the longitudinal direction) 
and $N_{\bm{e_{2}}}$ is the one along $\bm{e_{2}}$ (the transverse direction) and selecting periodic 
boundary conditions (PBC) in two directions, we plot the energy gap of the lower two minibands as a function of $V$ 
in Fig.~\ref{f2}(a). Intuitively, we can see that a critical point $V_{c}$ ($V_{c}\approx 0.82J$ in the numerical calculations) 
separates the system into two different phases. When the potential strength $V$ is less than the critical point, the gap 
is always open, whereas the gap is closed when $V$ crosses the critical point. To make it clear whether 
the energy gap is sensitive to the size of the system, 
we perform the finite-size analysis on the energy gap. Figures \ref{f2}(b1) and \ref{f2}(b2), \ref{f2}(c1) and \ref{f2}(c2), 
and \ref{f2}(d1) and \ref{f2}(d2) shows the results with $V=0.1J$, $V=0.5J$, and $V=J$, respectively. Particularly, 
to calculate the gaps in Figs.~\ref{f2}(b1), \ref{f2}(c1), and \ref{f2}(d1), we leave $N_{\bm{e_{2}}}$ unchanged. 
When calculating the gaps in \ref{f2}(b2), \ref{f2}(c2), and \ref{f2}(d2), we keep $N_{\bm{e_{1}}}$ invariant 
and make $N_{\bm{e_{2}}}$ be equal to the $s$-th Fibonacci number $F_s$. Obviously, as the size of the 
system increases, the gaps almost remain constants. It implies that the energy gap is insensitive to the size of 
the system. Based on the energy gap, we infers that the gapped phase at $V>0$ preserves the topological 
characteristics of the original uniform case ($V=0$). We know that for the uniform case, its topological 
properties are intermediately reflected from the Bloch Chern number. That is,  the TKNN formula \cite{TKNN} 
is used to characterize the topological properties of the system. In this case, the Chern 
number of the lowest bands will show to be $C_{1}=1$. 

 \begin{figure}[htp]
 \centering
 \includegraphics[width=0.5\textwidth]{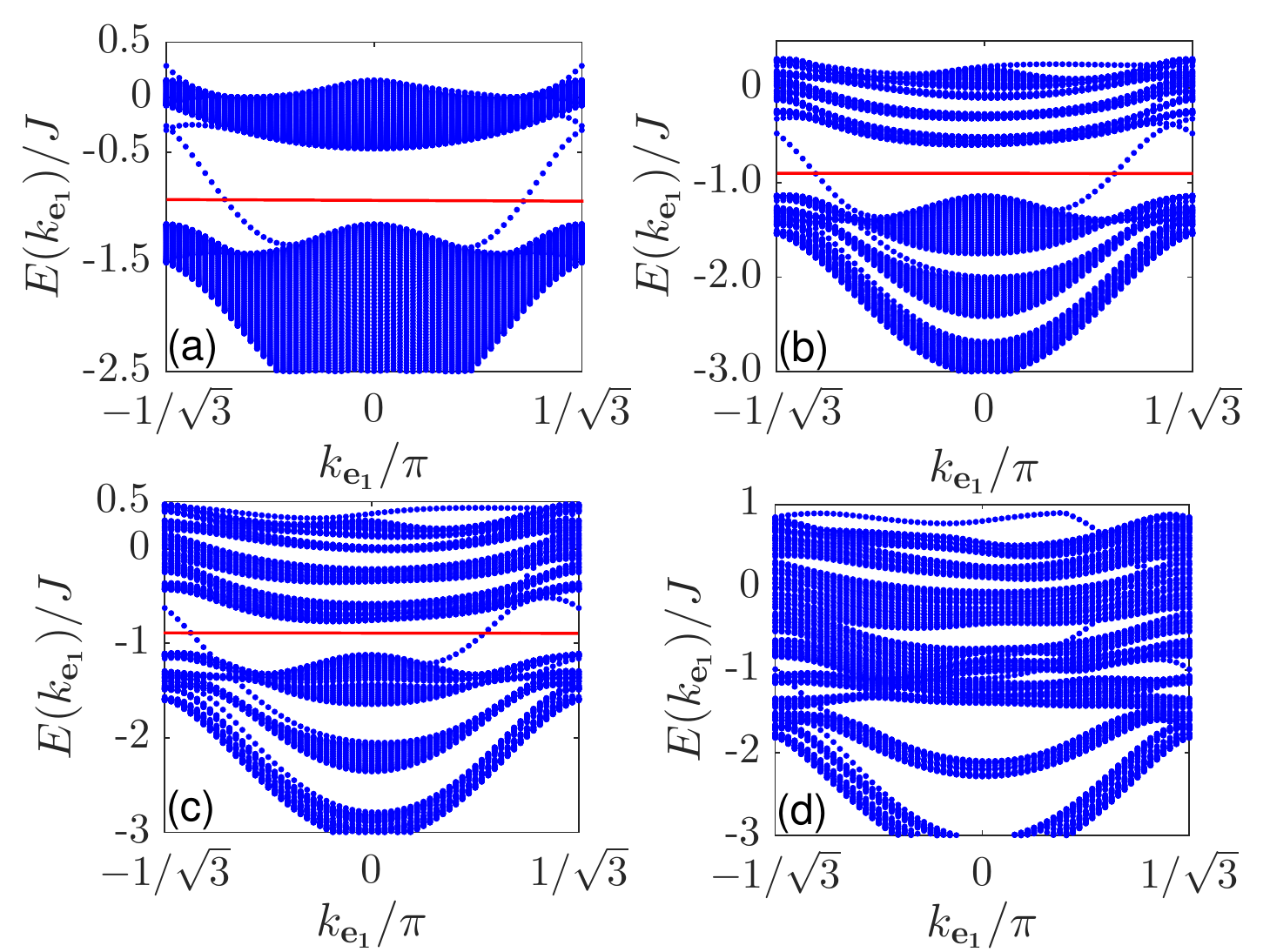}
 \caption{(Color Online) Zigzag-edge spectra $E(k_{\bm{e_{1}}})/J$ of the lowest two bands as a function 
 of $k_{\bm{e_{1}}}$. (a) $V=0J$; (b) $V=0.3J$; (c) $V=0.5J$; (d) $V=1J$. The red solid lines are the 
 representative Fermi levels. The size of the system is $64 \times 89$. 
 }\label{f3}
 \end{figure}

We first try to verify the aforementioned inference by the singly periodic spectrum. The lattice size is 
$64 \times 89$ and we leave it with PBC in the longitudinal direction and with zigzag edge in the transverse 
direction. Thus, the momentum $k_{\bm{e_{1}}}$ in the $\bm{e_{1}}$ direction is a good quantum number. 
Figure \ref{f3}(a) presents the zigzag-edge spectrum in the $V=0$ case. The red line represents the 
$1/3$-filling fermi energy level in the bulk gap, at which there are two edge modes with opposite momentum 
$k_{\bm{e_1}}$, whose corresponding group velocities are of opposite signs. For the quasi-periodic cases, one can 
see that there exists two edges within the bulk gap at $1/3$ filling as well (see the cases with $V=0.3J$ and 
$V=0.5J$ in Figs.~\ref{f3}(b) and \ref{f3}(c), respectively). It means that there are preserved topological 
characteristics in the quasi-periodic case. Nevertheless, compared to the uniform case, the corresponding 
momenta of the two edge modes are no longer symmetric about $k_{\bm{e_{1}}}=0$ in the quasi-periodic cases. 
For the $V=J$ case, there is no full bulk gap in the zigzag-edge spectrum (see Fig.~\ref{f3}(d)), presenting 
the metallic characteristics, which is self-consistent with the energy gap under PBC (see Fig.~\ref{f2}(a)).     

Except for the edge modes, the transport conductance \cite{TAI_1,TAI_2,TAI_theory,TAI_appli_1,TAI_appli_2,TAI_appli_3,mobility_gap} 
is an observable to characterize the topological features as well. 
For convenience but without loss of generality, both the (left and right) leads 
are described by the Hamiltonian $\hat{H}$ as well. The conductance $G_{E}$ of the 
system (central scattering region) can be obtained from the Landauer formula \cite{Lan_formula_1,Lan_formula_2}, 
\begin{equation}
G_{E}=\frac{2e^2}{h} T_{E},
\end{equation}
where $2e^2/h$ is the unit of $G_{E}$ and $T_{E}$ is the transmission coefficient, which is 
expressed as 
\begin{equation} 
T_{E}=Tr\left[\Gamma_{L} G^{r} \Gamma_{R} G^{a} \right], 
\end{equation}  
where $G^{r}$ ($G^{a}$) is the retarded (advanced) Green's function of the system with 
$G^{r}=\left[EI-H_{C}-\Sigma_{L}-\Sigma_{R}\right]$ and $G^{a}=\left(G^{r}\right)^{\dag}$, 
and $\Gamma_{L/R}=i\left[\Sigma_{L/R}-\left(\Sigma_{L/R}\right)^{\dag}\right]$. Here, we 
have used $H_{C}$ to denote the Hamiltonian of the system. The self-energies $\Sigma_{L}$ and 
$\Sigma_{R}$ are given by 
\begin{equation}
\Sigma_{L}=H^{\dag}_{LC}{\rm{g}}_{L}H_{LC}, ~\Sigma_{R}=H_{CR}{\rm{g}}_{R}H^{\dag}_{CR}, 
\end{equation}
where $H_{LC}$ ($H_{CR}$) denotes the coupling matrices between the system and the $L~(R)$ leads, and 
${\rm{g}}_{L}$ (${\rm{g}}_{R}$) are surface Green's functions of the $L~(R)$ leads. 

 \begin{figure}[htp]
 \centering
 \includegraphics[width=0.5\textwidth]{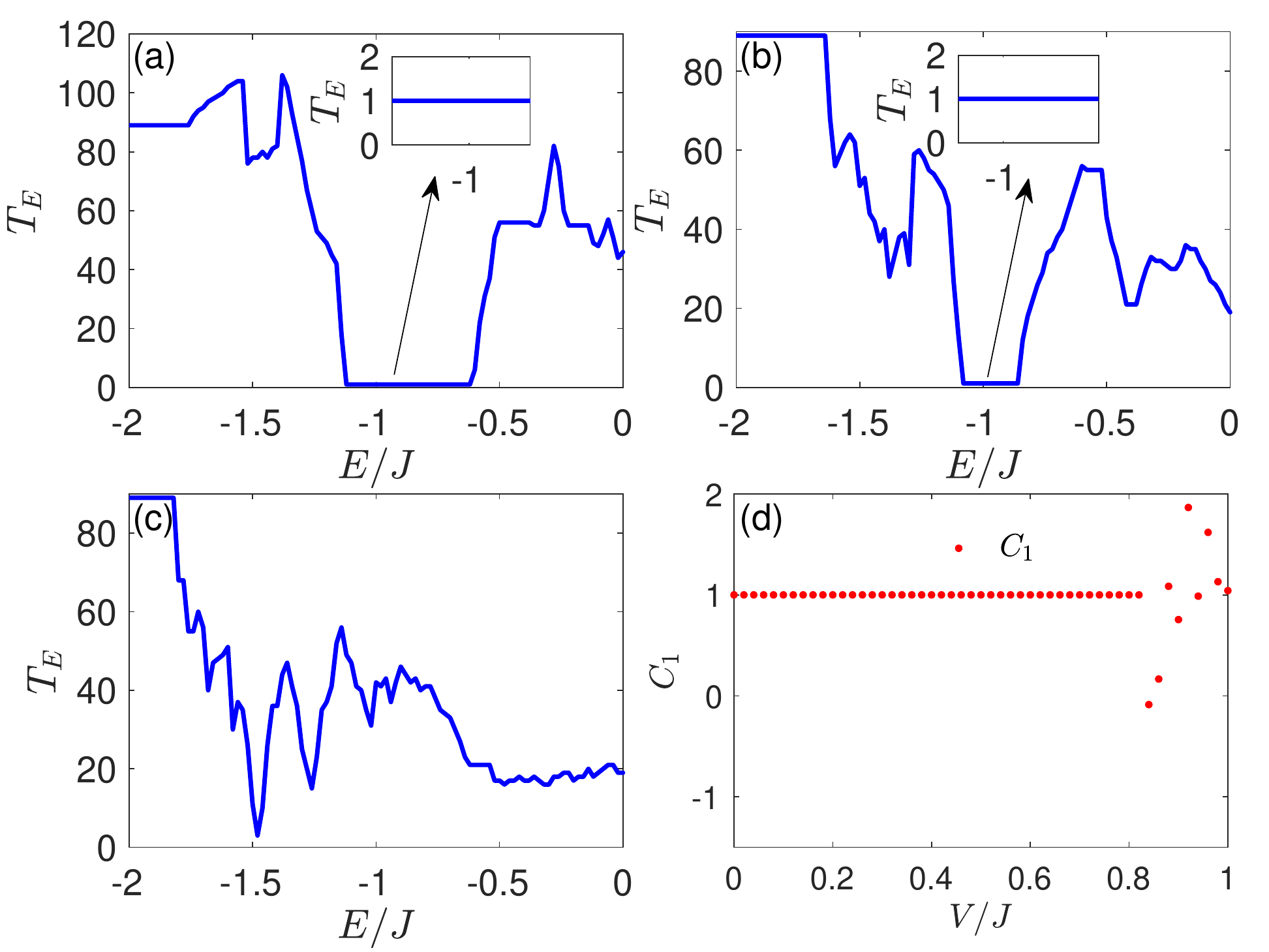}
 \caption{(Color Online) The transmission coefficient $T_E$ as a function of the Fermi energy with 
 $V=0.3J$ in (a); $V=0.6J$ in (b); and $V=J$ in (c). The insets enlarge the stage with $T_E=1$. 
 (d) The Chern number $C_{1}$ versus the potential strength $V$ at 1/3 filling. The size of the system 
 is $40 \times 144$. 
 }\label{f4}
 \end{figure}

We consider a system size $40 \times 144$ with OBC in $\bm{e_{2}}$ and take $V=0.3J$, $V=0.6J$, 
and $V=J$. The transmission coefficients as a function the Fermi energy are presented in Figs.~\ref{f4}(a), 
\ref{f4}(b), and \ref{f4}(c), respectively. Apparently, in the gapped cases ($V=0.3J$ and $V=0.6J$), there 
is a step with $T_E=1$ at 1/3 filling. While in the gapless case ($V=J$), no stage with $T_{E}=1$ exists 
at 1/3 filling. Furthermore, the results of the $T_{E}$ are coincide with the topological phase diagram in Fig.~\ref{f4}(d).  
In this diagram, $C_{1}=1$ for $V < V_{c}$, corresponding to $T_{E}=1$,  while $C_{1}$ is unquantized 
when the potential strength $V$ crosses the critical point $V_{c}$, corresponding to the unquantized $T_{E}$. 
The reason for the appearance of the unquantized Chern number is that in this parameter region, the bulk gap 
is actually closed. Hence, there is no topology-protected edge mode, reflecting the trivial nature of the system. 
On the contrary, the quantized $C_{1}=1$ exactly corresponds to the two edge modes within the bulk gap, 
presenting the bulk-edge correspondence, and being self-consistent with the two-channel conductance. 
In addition, the system can be viewed as a one-dimensional tight-binding model in the longitudinal direction, 
but each unit cell now is replaced by a super cell which contains $N_{\bm{e_2}}$ cells. When $V>V_{c}$, 
the one-dimensional tight-binding model describes a gapless metal. Therefore, in the gapless case, 
there is no quantized conductance below or above the Fermi energy at 1/3 filling (see Fig.~\ref{f4}(c)), presenting 
intrinsically metallic transport characteristics of the system.  

\paragraph{Localization phase transition.---} The two-channel transport observation shows that 
there is no any edge mode and no localization phenomenon along the longitudinal direction when $V>V_{c}$. 
We infer that the absence of the edge mode is related with the localization in the transverse
direction. After all, the localization will prevent the particles from moving towards the boundaries of 
the system, so that the system can not form the edge states. To confirm this conjecture, we will use 
the partial inverse participation ratio (PIPR) to characterize the localization properties of the system. 
The system has a size with $N_{\bm{e_{2}}}=F_{s}$ in the $\bm{e_{2}}$ direction while it has no limitation
for $N_{\bm{e_{1}}}$, but satisfies PBC in two directions. $N_{\bm{e_{2}}}$ is chosen as the Fibonacci 
number to minimize the size effect. Then, the PIPR of a normalized 
wave function $\psi$ reads 
\begin{equation}\label{PIPR}
{\rm PIRP}=\sum^{N_{\bm{e_{2}}}-1}_{n=0} | \psi({\mathcal{S}_{n}}) |^4N^2_{\bm{e_{1}}}.
\end{equation}
where the index $m$ has been suppressed (the same below).

 \begin{figure}[htp]
 \centering
 \includegraphics[width=0.5\textwidth]{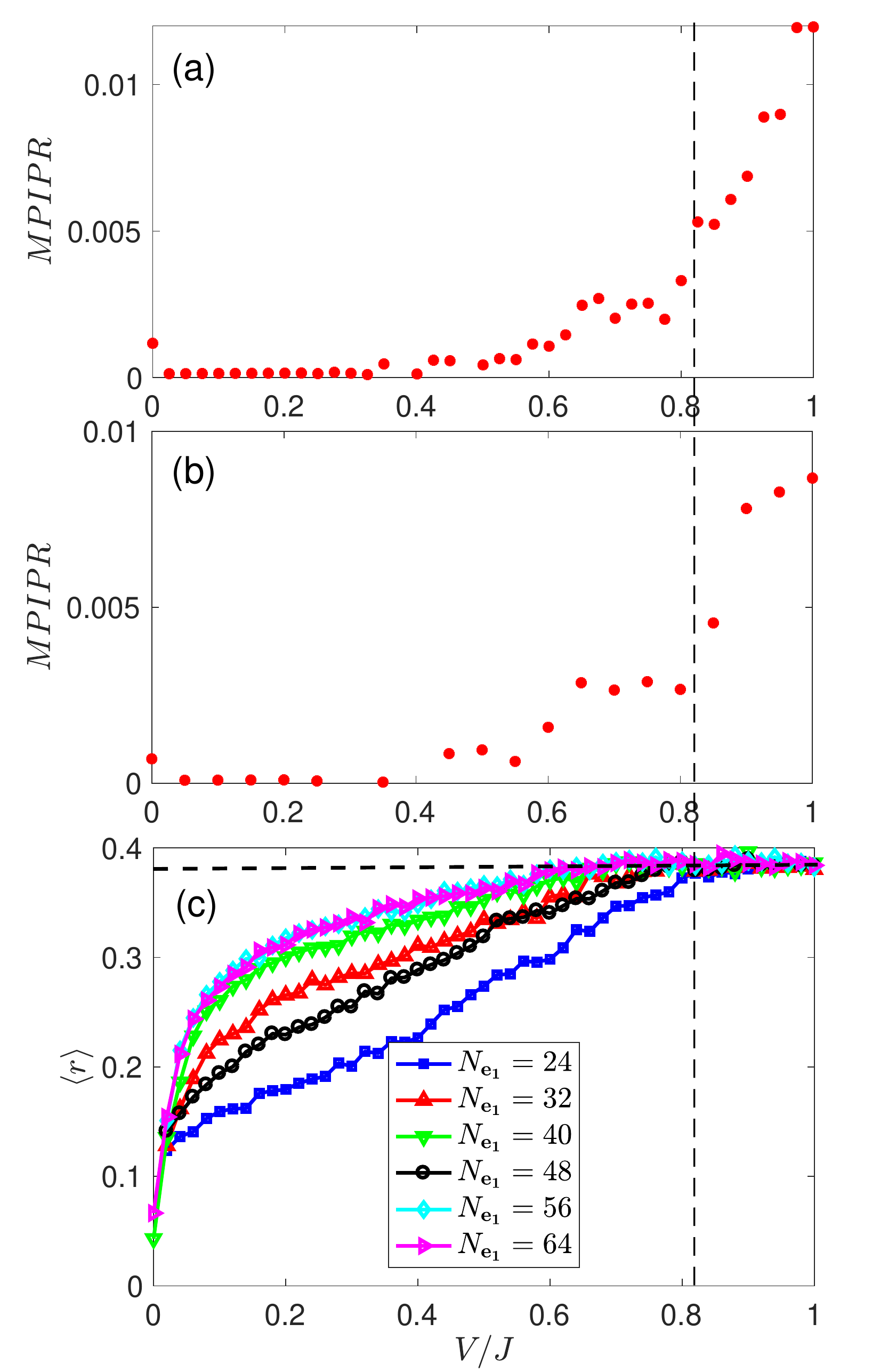}
 \caption{(Color Online) The MPIPR as function of the quasi-periodic potential strength $V$ with the system 
 size $24\times 233$ in (a) and $24\times 377$ in (b). The averaged level spacing ratio $\langle r \rangle$ 
 as a function of $V$ is presented in (c) with $N_{\bf e_{2}}=144$, where the horizontal black dashed line corresponds to the Poisson 
 distribution value $\langle r \rangle \approx 0.3863$. The black dashed reference 
 line denotes the critical point.}\label{f5}
 \end{figure}

Since there is preserved translational invariance in the longitudinal direction, the wave packets is 
periodically distributed in this direction. Noting that the PIPR in Eq.~(\ref{PIPR}) is a 
quantity defined in the transverse direction, hence, it is more convenient and effective to use 
the PIPR to characterize the localization property in this direction. When the PIRP tends 
to a finite value $O(1)$, the wave function is localized in the $\bm{e_{2}}$ direction. The 
PIRP scales as $F^{-1}_{s}$ for the extended states, while behaves like $F^{-\gamma}_{s}$ ($0 < \gamma < 1$)
for the critical states. Due to the distinct difference between the localized states and the extended 
(or critical) states, therefore, in the following, we name the extended and the critical states as the 
delocalized states. Choosing two system sizes $24\times 233$ and $24 \times 377$ as examples, the corresponding 
mean PIPR (MPIPR) of the wave functions within the 1/3 filling are plotted in Figs.~\ref{f5}(a) and 
\ref{f5}(b). Intuitively,  the MPIPR jumps at the critical point (the black dashed line shows), signaling a 
transverse delocalization-localization transition.

Next, we try to study the transverse localization transition by analyzing the energy gap statistic. 
To perform the analysis, we will calculate the average of the energy level spacing ratio $\langle r \rangle$ over 
the energies within the 1/3 filling, which is defined by 
$\langle r \rangle=\frac{1}{N_{\bm{e_{1}}}\times N_{\bm{e_{2}}}-2}\sum_{j} r_{j}$, 
where $r_{j}={\rm min}\{\delta_{j},\delta_{j+1}\}/{\rm max}\{\delta_{j},\delta_{j+1}\}$ and $\delta_{j}=E_{j+1}-E_{j}$ 
with the energies $E_{j}$ arranged in an ascending order. In the numerically calculations, we take PBC and $N_{\bf e_{2}}=144$. 
Figure.~\ref{f5}(c) presents $\langle r \rangle$ as a function of $V$ for various $N_{\bf e_{1}}$ at 1/3 
filling. Apparently, $\langle r \rangle$ approaches the Poisson distribution value \cite{poisson_1,poisson_2} 
$\langle r \rangle_{\rm Possion}=2{\rm ln}(2)-1\approx 0.3863$ (the horizontal black dashed line in Fig.~\ref{f5}(c) shows) 
when the quasidisorder strength $V$ is larger than the critical point. Besides, it is readily seen that $\langle r \rangle$ 
for various $N_{\bf e_{1}}$ are less than $0.3863$ when $V$ is smaller than the critical value, presenting the level statistics 
feature of the single-particle delocalized states \cite{Huhui}.

 \begin{figure}[htp]
 \centering
 \includegraphics[width=0.5\textwidth]{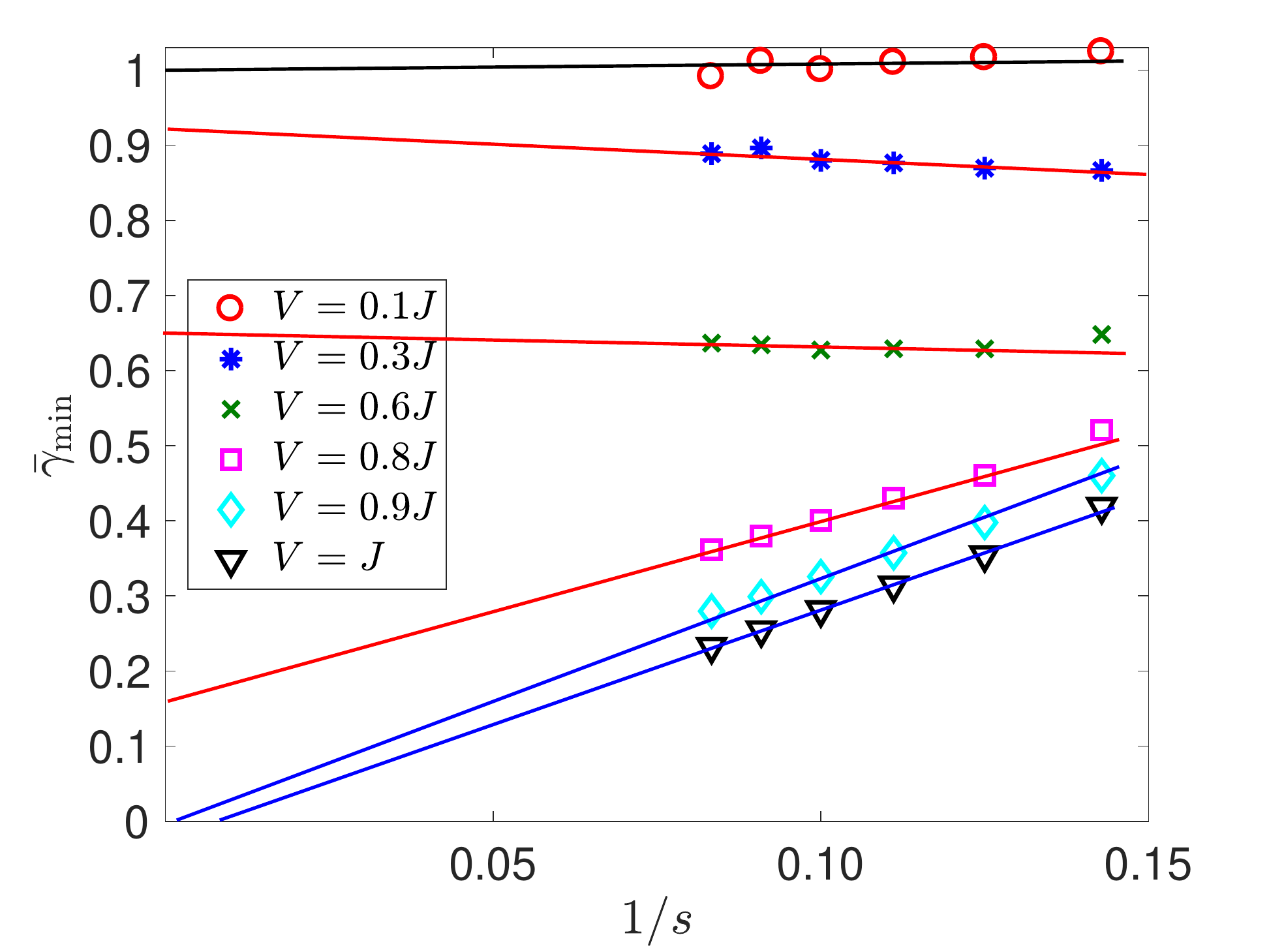}
 \caption{(Color Online) Plots of $\bar{\gamma}_{\rm min}$ versus $1/s$ for the wave functions within 1/3 filling.}\label{f6}
 \end{figure}

Finally, to make the transverse delocalization-localization transition more clear, we analyze the scaling 
behavior of the wave functions. As mentioned above, the wave packets are periodically distributed at each site 
$\mathcal{S}_{n}$. The probability $P(\mathcal{S}_{n})$ for the sites in the transverse direction is defined 
as $P(\mathcal{S}_{n})=|\psi(\mathcal{S}_{n})|^2 N_{\bm{e_{1}}}$. 
Then a scaling index $\gamma_{j}$ is given by $P(\mathcal{S}_{n})=F^{-\gamma_{j}}_{s}$. For a concrete wave function, 
$\gamma_{j}$ will distribute within a interval $\left[\gamma_{\rm min},~\gamma_{\rm max}\right]$. In the thermodynamic 
limit $1/s\rightarrow 0$, $\gamma_{\rm min}\rightarrow 0$ corresponds to the localized states, 
while $0 < \gamma_{\rm min} \leq 1$ for the delocalized states (Particularly, $\gamma_{\rm min}=1$ denotes the 
extended states, otherwise, it corresponds to the critical states). Therefore, we will employ $\gamma_{\rm min}$ 
to characterize the scaling behaviors of the wave functions. In the numerical calculations, we choose $N_{\bm{e_{1}}}=40$ 
and take PBC and then average the $\gamma_{\rm min}$ over the wave functions whose corresponding energies 
are within the $1/3$ filling. We label the averaged $\gamma_{\rm min}$ by  
$\bar{\gamma}_{\rm min}=\frac{1}{N_{\bm{e_{1}}}\times N_{\bm{e_{2}}}}\sum^{N_{\bm{e_{1}}}\times N_{\bm{e_{2}}}}_{j=1}\gamma^{j}_{\rm min}$.    
The extracted $\bar{\gamma}_{\rm min}$ from the wave functions for different $V$ are plotted in Fig.~\ref{f6}. 
From this plots, it is readily seen that the $\bar{\gamma}_{\rm min}$ in the extrapolating limit decreases as the quasi-periodic 
potential strength $V$ increases, and it finally reduces to zero when $V$ crosses the critical point $V_{c}$, verifying the 
transverse delocalization-localization phase transition. The results reflected from the $\bar{\gamma}_{\rm min}$ confirms 
the prediction of the MPIPR and the energy gap statistics.

\paragraph{Summary.---} In summary, we have investigated the influence of the quasidisorder on the topological properties 
and the localization behaviors of a two-dimensional system. The topological phase transition is charactered by the band gap, 
edge-state spectra, and transport conductance. When the topological transition happens, we find that the bands gap closes, 
and no any edge modes exists. Meanwhile, the transport conductance shows that the system is more like a metal 
in the gapless phase. The results are self-consistent with the topological diagram which contains the Chern number 
of lowest band. In addition, a transverse localization transition is characterized by the MPIPR, level 
statistics, and the fraction dimension. The findings extend the insight about relationship between the topology and the 
localization properties of the bulk states, and the reveal the topological nature of the bulk delocalized states.

In the previous work, we have made a theoretical scheme \cite{Shujie} to realize the Hamiltonian $\hat{H}_{1}$ presented in Eq.~(\ref{eq1}). 
In this scheme, the noninteracting particles are confined in the two-dimensional lattice, and initially hop between the 
nearest-neighbor sites. The system is initially gapless and topological trivial. Once a circular-frequency shaking is applied, 
then the next-nearest hoppings are introduced and then the system becomes gapped and topological nontrivial. Experimentally,  
the lattice geometry has been realized by three retro-reflected lasers and the hoppings of particles are controlled by 
tuning the depth of the \cite{Eckardt}. Besides, the Floquet shaking technique has been applied to optical 
honeycomb lattice \cite{Esslinger} to realize the Haldane model \cite{Chern}. In addition, the quasidisorder potential $\hat{H}_{2}$ 
presented in Eq. (\ref{eq2}) has been realized by superimposing two independent lasers with different 
wavelengths \cite{wave_1,wave_2,wave_3}. As a result, we believe that the findings in this two-dimensional quasidisordered 
system has the potential to be observed in the ultracold atomic experiments.

We acknowledge support from NSFC under Grants No. 11835011 and No. 12174346.

\bibliography{references}
\end{document}